\documentclass[prd,showpacs,aps,preprint]{revtex4}
\usepackage{graphicx}
\usepackage{dcolumn}
\usepackage{amsfonts}
\usepackage{amsmath}

\newcommand{\ep}{\epsilon}

\newcommand{\pa}{\partial}

\newcommand{\beq}[1]{\begin{eqnarray}\label{#1}}
\newcommand{\eeq}{\end{eqnarray}}

\newcommand{\nn}{\nonumber}

\begin{document}
\title{Cosmological Constant as Vacuum Energy Density of Quantum Field
Theories on Noncommutative Spacetime}
\author{Xiao-Jun Wang}
\email{wangxj@ustc.edu.cn}
\affiliation{\centerline{Interdisciplinary Center for Theoretical
Study} \centerline{University of Science and Technology of China}
\centerline{AnHui, HeFei 230026, China}}

\begin{abstract}
We propose a new approach to understand hierarchy problem for
cosmological constant in terms of considering noncommutative
nature of space-time. We calculate that vacuum energy density of
the noncommutative quantum field theories in nontrivial
background, which admits a smaller cosmological constant by
introducing an higher noncommutative scale $\mu_{_{NC}}\sim M_p$.
The result $\rho_\Lambda\sim
10^{-6}\Lambda_{SUSY}^8M_p^4/\mu_{_{NC}}^8$ yields cosmological
constant at the order of current observed value for supersymmetry
breaking scale at $10$TeV. It is the same as Banks'
phenomenological formula for cosmological constant.
\end{abstract}
 \preprint{USTC-ICTS-04-23} \maketitle

It is great puzzle to understand the hierarchy problem on current
observed tiny cosmological constant, $\rho_\Lambda\sim
10^{-12}$eV$^4$. The puzzle may be stated by identification
between cosmological constant and vacuum energy density of quantum
field theories (QFT),
\beq{1}\rho_\Lambda\sim \int
d^3\vec{k}\;\omega_{\vec{k}}\sim
\int_0^{\Lambda^2}dk^2\;k^2=\Lambda^4.
\eeq
For $\Lambda=M_p$ with reduced Plank scale $M_p=(8\pi
G)^{-1/2}\simeq 2.4\times 10^{18}$GeV it exhibits huge hierarchy
of $M_p^4/\rho_\Lambda\sim 10^{120}$. Usually one may introduce
supersymmetry to improve this hierarchy problem while the
identification between cosmological constant and vacuum energy of
QFT is insisted. In this scenario the contributions to vacuum
energy from bosons and fermions are exactly cancelled each other.
Unfortunately, current particle physics experiments reject the
supersymmetry at scale below TeV. Then naive treatment
$\rho_\Lambda=\Lambda_{SUSY}^4$ with $\Lambda_{SUSY}$ the scale of
supersymmetry breaking still produces too large cosmological
constant. However, to discuss the cosmological constant problem,
we need to bring gravity into the picture. Banks\cite{Banks01},
therefore, suggested a phenomenological formula to link the scale
of supersymmetry breaking and the Plank mass to the cosmological
constant,
\beq{1a}\rho_\Lambda\;\sim \;\Lambda_{SUSY}^4
\left(\frac{\Lambda_{SUSY}}{M_p}\right)^4.
\eeq
This formula yields the cosmological constant at the order of
current observed value for $\Lambda_{SUSY}$ at TeV. So far,
however, it is not well-understood what is theoretical origination
of (\ref{1a}). Alternatively one may further introduce a mechanism
which greatly suppresses the contribution from high energy in
Eq.~(\ref{1}). This mechanism is expected to somehow exhibits some
basic properties of quantum gravity or of spacetime. The purpose
of this letter is just to propose one of such kind of mechanism by
studying that vacuum energy density of QFT at noncommutative
spacetime (NCQFT),
\beq{2}[x^\mu,\;x^\nu]=\ep^{\mu\nu}/\mu_{_{NC}}^2,
\eeq
where $\mu_{_{NC}}$ is a scale to specify the noncommutative
effect and the components of anti-symmetric tensor $\ep^{\mu\nu}$
takes the values $1$, $-1$ or $0$. We shall show that this
mechanism together with supersymmetry predicts the
formula~(\ref{1a}).

The action for field theories on noncommutative spaces can be
obtained using the Weyl-Moyal correspondence~\cite{BFFL78},
according to which, a traditional treatment is to replace the
usual product in field theory by the so-called
star-product~\cite{SW99,MM00}. In this present letter, however, we
will use another more convenient realization of algebra~(\ref{1})
by reordering operators in the plane wave basis\cite{Douglas01}
\beq{3}e^{iq\cdot x}e^{ik\cdot x}=e^{-\frac{i}{2}q\times k}
e^{i(q+k)\cdot x},
\eeq
where
\beq{4}q\times k=-k\times q=\frac{1}{\mu_{_{NC}}^2}\ep^{\mu\nu}q_\mu
k_\nu.
\eeq
Following Eq.~(\ref{3}) we have the properties
\beq{5}\int d^4x e^{iq\cdot x}e^{ik\cdot x}&=&\delta^{(4)}(q+k),
\nn \\
\int d^4x e^{iq\cdot x}e^{ik\cdot x}e^{ip\cdot x} &=&\int d^4x
e^{ip\cdot x}e^{iq\cdot x}e^{ik\cdot x}.
\eeq
It has been shown that the most of NCQFTs are renormalizable at
one-loop level at least\cite{MS99,MM001}. In this present letter
we consider the simplest model that describes a free scalar
particle$^{1}$ \footnotetext[1]{Our discussions can be extended to
free fields with diverse spin. The results will be same besides
that an extra factor $-1$ is contributed by fermions.}moving in
four-dimensional noncommutative curved spacetime. Its action is
specified by
\beq{6}S=\frac{1}{2}\int d^4x\sqrt{-g} [g^{\mu\nu}
\pa_\mu\phi\pa_\nu\phi+m^2\phi \phi]+S_{EH},
\eeq
with $S_{EH}$ the standard Einstein-Hilbert action. For sake of
convenience we will work in Euclidean signature via substitution
$x^0\to ix^4$ and set that
$\ep^{12}=\ep^{34}=-\ep^{21}=-\ep^{43}=1$ while other components
of $\ep^{\mu\nu}$ are set to zero.

The generating functional is obtained by path integral over
$\phi$:
\beq{7}Z[g,m^2]={\cal N}\int [d\phi(x)]e^{-S}
={\cal N}\exp{\{-\frac{1}{2}{\rm
Tr}\ln{(-\pa_\mu(\sqrt{g}g^{\mu\nu}\pa_\nu+m^2\sqrt{g})}+S_{EH}\}},
\eeq
where ${\cal N}$ is an infinite constant from path integral over
of canonical momentum of $\phi$ and ``Tr'' is a trace over
noncommutative space-time. To be precise, Eq.~(\ref{7}) can be
rewritten as follows by using Eq.~(\ref{3}) and (\ref{5}),
\beq{8}
\ln{Z[g,m^2]}-S_{EH}&=&-\frac{1}{2}\int
d^4x\int\frac{d^4k}{(2\pi)^4}
\sqrt{g(x)}e^{-ik\cdot x}{\cal D}(x;k)e^{ik\cdot x} \nn \\
&=&-\frac{1}{2}\int d^4x\frac{d^4y}{(2\pi)^4}
\frac{d^4w}{(2\pi)^4} \int\frac{d^4k}{(2\pi)^4}
\sqrt{g(w)}\delta^{(4)}(w-x) \nn \\
&&\quad\quad\quad\cdot e^{-ik\cdot x}\delta^{(4)}(y-x)
e^{ik\cdot x}{\cal D}(y;k) \nn \\
&=&-\frac{1}{2}\int d^4w\frac{d^4y}{(2\pi)^4}
\int\frac{d^4k}{(2\pi)^4}\frac{d^4q}{(2\pi)^4}
\sqrt{g(w)}e^{iq\cdot (w-y)}e^{-iq\times k}{\cal D}(y;k),
\eeq
where ${\cal D}(x;k)=\ln{\{[\sqrt{g(x)}(g^{\mu\nu}(x)k_\mu
k_\nu+m^2)+\pa_\mu(\sqrt{g(x)}g^{\mu\nu}(x))k_\nu]/\Lambda^2\}}$
with $\Lambda$ the intrinsic UV cut-off of QFT which is defined by
${\cal N}=\exp{\{\frac{1}{2}\ln{\Lambda^2}\;{\rm Tr {\bf I}}\}}$
with ${\bf I}$ identity operator. In Eq.~(\ref{8}) we use the
plane wave expansion, $\langle k|x\rangle=e^{ik\cdot x}$ and
$\delta^{(4)}(x-y)\sim \int d^4q\; e^{iq\cdot (x-y)}$. It is
valid, as the argument below, in a region whose size is much
smaller than cosmological event horizon since our current observed
universe is rather flat.

The ``true'' vacuum amplitude should be obtained by further
integral over metric field $g$, i.e., the degrees of freedom of
quantum gravity. So far, unfortunately, we do not know how to
construct a consistent theory of quantum gravity in four
dimensions and how to perform the integral over its degrees of
freedom. Alternatively, we may assume that the spacetime geometry
is described by classical solution of Einstein equation at the
distance much larger than Plank length, i.e., $|x-y|\gg 1/M_p$ or
$q^2\ll M_p^2$. For example, we can use the de Sitter (static)
metric with Euclidean rotation,
\beq{8a}
ds^2=\left(1-\frac{r^2}{L^2}\right)d\tau^2
+\left(1-\frac{r^2}{L^2}\right)^{-1}dr^2+r^2d\Omega_2^2.
\eeq
The spacetime described by the above metric is very flat in a
region whose size is much smaller than the event horizon $L$.
Hence Eq.~(\ref{8}) may be split into two parts,
\beq{9}\ln{Z[g,m^2]}&=&-\frac{1}{2}\int_L d^4x\frac{d^4y}{(2\pi)^4}
\int_{0}^{M_p} \frac{|q|^3d|q|}{8\pi^2} \int\frac{d^4k}{(2\pi)^4}
e^{iq\cdot (x-y)}e^{-iq\times k}\ln{\frac{k^2+m^2}{\Lambda^2}} \nn \\
&&+\Gamma_s[g,m,M_p],
\eeq
where $\int_L d^4y$ denotes to cut integral over $y$ at the event
horizon and $\Gamma_s[g,m,M_p]$ specifies the short-distance
contribution which contains the degrees of freedom of quantum
gravity. The UV cut-off at Plank scale in integral over $q$ is
significant. It partly reflects effects of quantum gravity, and
should be distinguish with cut-off $\Lambda$, which by definition
is characteristic scale of QFT, e.g., the supersymmetry breaking
scale $\Lambda_{SUSY}$. Then the vacuum amplitude is explicitly
specified by
\beq{10}Z_{vac}[m^2,M_p]=\int [dg(x)]Z[g,m^2].
\eeq
If we ignore the contribution from quantum gravity, the vacuum
energy density from NCQFT reads off$^2$ \footnotetext[2]{Precisely
we should further add an IR cut-off in integral over $q$, i.e.,
$q^2\geq H^2$ with $H$ the Hubble constant, by incorporating the
fact that the size of our universe is finite. Since result is
proportional to $M_p^2-H^2$ we ignore the Hubble constant in the
result.}
\beq{11}\rho_0&=&-\frac{1}{(4\pi)^2}\int_L\frac{d^4y}{(2\pi)^4}
\int_{0}^{M_p} |q|^3d|q|
\int\frac{d^4k}{(2\pi)^4} e^{iq\cdot y}e^{-iq\times
k}\ln{\frac{k^2+m^2}{\Lambda^2}} \nn
\\ &\to & \int_L\frac{d^4y}{(2\pi)^4}
\int_{0}^{M_p} \frac{|q|^3d|q|}{(4\pi)^2} e^{iq\cdot y}
\int\frac{d^4k}{(2\pi)^4} e^{-iq\times k}\int_0^\infty\frac{dl}{l}
\exp{\{-l(k^2+m^2)-\frac{1}{l\Lambda^2}\}}.
\eeq
To integrate over the $k$ we obtain
\beq{12}\rho_0&=&\int_L\frac{d^4y}{(2\pi)^4}
\int_{0}^{M_p} \frac{|q|^3d|q|}{(4\pi)^2} \int\frac{dl}{16\pi^2 l} \nn \\
&&\times
\exp{\left\{-l(\frac{q^2}{4\mu_{_{NC}}^4}+\frac{1}{\Lambda^2})
-\frac{m^2}{l}+iq\cdot y\right\}}.
\eeq
The factor $\frac{q^2}{4\mu_{_{NC}}^4}+\frac{1}{\Lambda^2}$ is
known to denote the UV/IR-mixing phenomenon, which is typical
properties of NCQFT\cite{Mw2000}. Finally to perform all integrals
in Eq.~(\ref{12}) we get:
\beq{13}\rho_0&= &\frac{\mu_{_{NC}}^8}{32\pi^4}
\int_0^\infty\frac{dl}{l} \int_L\frac{d^4y}{(2\pi)^4}
\left[1-(1+lM_p^2)e^{-lM_p^2}\right]
\exp{\left\{-\frac{4\mu_{_{NC}}^4}{\Lambda^2}l
-\frac{m^2}{4l\mu_{_{NC}}^4}-\frac{y^2}{4l}\right\}} \nn \\
&\simeq & \frac{\Lambda^4}{512\pi^6}\int_0^\infty dl\;l
\left[1-(1+l\kappa)e^{-l\kappa }\right]
\exp{\left\{-l-\frac{m^2}{l\Lambda^2}\right\}}
\nn \\
&=&\frac{m^2\Lambda^2}{256\pi^6}\left(K_2(\frac{2m}{\Lambda})-
\frac{1}{1+\kappa}K_2(\frac{2m}{\Lambda}\sqrt{1+\kappa})
-\frac{m}{\Lambda}\frac{\kappa}{(1+\kappa)^{3/2}}
K_3(\frac{2m}{\Lambda}\sqrt{1+\kappa})\right),
\eeq
where $K_2,\;K_3$ are modified Bessel function of the second kind
and $\kappa=\Lambda^2M_p^2/(4\mu_{_{NC}}^4)$. For $\kappa \gtrsim
1$ we just recover the result of usual QFT, $\rho_0\sim \Lambda^4$
for arbitrary value of $m\leq \Lambda$. For $\kappa\ll 1$,
however, from the second line of Eq.~(\ref{13}) we have
\beq{14}\rho_0&\simeq & \frac{\Lambda^4}{512\pi^6}\int_0^\infty dl\;l
\left[1-(1+l\kappa)(1-l\kappa+\frac{l^2\kappa^2}{2}+{\cal
O}(\kappa^3))\right]
\exp{\left\{-l-\frac{m^2}{l\Lambda^2}\right\}} \nn \\ &\simeq&
\frac{\kappa^2\Lambda^4}{1024\pi^6}\int_0^\infty dl\;l^3
\exp{\left\{-l-\frac{m^2}{l\Lambda^2}\right\}}+{\cal O}(\kappa^3)
\nn \\ &\simeq & \frac{6\Lambda^8
M_p^4}{2^{14}\pi^6\mu_{_{NC}}^8}\left(1-\frac{m^2}{3\Lambda^2}+{\cal
O}(\frac{m^4}{\Lambda^4})\right)+{\cal O}(\kappa^3).
\eeq
If we believe the noncommutativity is intrinsic property of
spacetime, we should expect $\mu_{_{NC}}\sim M_p$. Then for
$\Lambda=M_p$ we obtain usual result $\rho_\Lambda\sim M_p^4$.
However, if we expect our world is supersymmetric at higher
energy, one has $\Lambda\sim\Lambda_{SUSY}$ such that
\beq{15}\rho_\Lambda\sim 10^{-6}\frac{\Lambda_{SUSY}^8}{M_p^4}.
\eeq
This is precise formula~(\ref{1a}) of Banks' phenomenological
hypothesis on cosmological constant \cite{Banks01,Banks04}. For
lower supersymmetry breaking scale at $10$TeV, Eq.~(\ref{15})
reproduces current observed value of cosmological constant.

Naively the results~(\ref{14}) and (\ref{15}) should be doubted
since our traditional experience tell us that the physics inside
very short distance should not exhibit observable effects at low
energy. In other words, the results deduced from NCQFT should
return to one of commutative QFT at the limit
$\mu_{_{NC}}\to\infty$. However, it is not case of our result.
This phenomenon is well-known in NCQFT that $\mu_{_{NC}}\to\infty$
is not always a smooth limit to commutative case due to the
presence of UV/IR-mixing\cite{Mw2000,Douglas01}. Technically, it
is basic property of so-called non-planar diagrams in NCQFT while
planar diagrams reduce to usual results of commutative QFT. In our
treatment of Eq.~(\ref{8}), however, we can not consistently
introduce ``planar''-like terms when metric as external source is
presented. In other words, it is by definition
``non-planar''-like. The basic physics behind non-planar diagrams
is that they exhibit non-local properties of NCQFT, in which the
physics inside the very short-distance produces observable effects
at long-distance. Therefore, the crucial point hidden in our
calculations is the UV/IR-mixing.

In conclusion, in this letter we suggested that the
noncommutativity of spacetime may suppress vacuum energy density
of QFT to understand the cosmological constant problem. We studied
the model describing free particles in the noncommutative
spacetime. In our treatment, the spacetime noncommutativity as
well as quantum gravity play significant roles. Although the later
can not be dealt with in detail, it introduces a characteristic
scale, the Plank scale, into our model. It implies that our model
has partly included the effect of quantum gravity. It is rather
interesting our model reproduces Banks' phenomenological formula
for cosmological constant. For supersymmetry breaking scale at
$10$TeV, we obtain current observed value of cosmology constant.

It is pleasure to thank Y.-F. Chen and J.-X. Lu for useful
discussions, and thank the organizers of ShangHai Workshop on
string/M theory, where this work was inspired. This work is partly
supported by the NSF of China, Grant No. 10305017, and through
USTC ICTS by grants from the Chinese Academy of Science and a
grant from NSFC of China.

\end{document}